\documentclass[11pt]{article}

\usepackage{arxiv}

\usepackage[utf8]{inputenc} 
\usepackage[T1]{fontenc}    
\usepackage{hyperref}       
\usepackage{url}            
\usepackage{booktabs}       
\usepackage{amsfonts}       
\usepackage{amsmath}
\usepackage{amssymb}
\usepackage{nicefrac}       
\usepackage{microtype}      
\usepackage{lipsum}		
\usepackage{graphicx}
\usepackage{natbib}
\usepackage{doi}
\usepackage{xcolor}
\usepackage{listings}
\definecolor{commentgreen}{RGB}{2,112,10}
\definecolor{codegreen}{rgb}{0,0.6,0}
\definecolor{codegray}{rgb}{0.5,0.5,0.5}
\definecolor{codepurple}{rgb}{0.58,0,0.82}
\definecolor{backcolour}{rgb}{0.95,0.95,0.92}

\lstdefinelanguage{ASM}{
    morekeywords={sin, time, rateOf, uniform, normal,
       maximize, constraint, model, end, notes},
    sensitive=false, 
    morecomment=[l]{//}, 
    morecomment=[s]{/*}{*/}, 
    morestring=[b]" 
} %

\lstdefinestyle{mystyle}{
    language=python,
    commentstyle=\color{codegreen},
    keywordstyle=\color{magenta},
    numberstyle=\tiny\color{codegray},
    stringstyle=\color{codepurple},
    basicstyle=\ttfamily,
    breakatwhitespace=false,         
    breaklines=true,                 
    captionpos=b,                    
    keepspaces=true,  
    frame=single,               
    numbersep=5pt,                  
    showspaces=false,                
    showstringspaces=false,
    showtabs=false, 
    columns=fullflexible,
    tabsize=2
}

\title{Computing the Frequency Response of Biochemical Networks: A Python module}

\author{\href{https://orcid.org/0000-0002-3659-6817}{\includegraphics[scale=0.06]{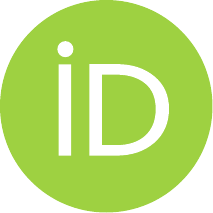}\hspace{1mm}Herbert M.~Sauro}\\
	Department of Bioengineering\\
	University of Washington\\
	Seattle, WA 98195-5061 \\
	\texttt{hsauro@uw.edu} } 

\renewcommand{\shorttitle}{Frequency Response}

\hypersetup{
pdftitle={Frequency Response},
pdfsubject={systems biology},
pdfauthor={Herbert M.~Sauro},
pdfkeywords={Antimony, SBML, Systems Biology},
colorlinks,
linkcolor=blue,
citecolor=blue,
urlcolor=blue
}

\begin{document}
\maketitle

\begin{abstract}
In this paper, a set of Python methods is described that can be used to compute the frequency response of an arbitrary biochemical network given any input and output. Models can be provided in standard SBML or Antimony format. The code takes into account any conserved moieties so that this software can be used to also study signaling networks where moiety cycles are common. A utility method is also provided to make it easy to plot standard Bode plots from the generated results. The code also takes into account the possibility that the phase shift could exceed 180 degrees which can result in ugly discontinuities in the Bode plot. In the paper, some of the theory behind the method is provided as well as some commentary on the code and several illustrative examples to show the code in operation. Illustrative examples include linear reaction chains of varying lengths and the effect of negative feedback on the frequency response.  

Software License: MIT Open Source

Availability: The code is available from \url{https://github.com/sys-bio/frequencyResponse}.
\end{abstract}


\keywords{Frequency Response \and SBML \and Systems Biology \and Python}

\section{Introduction}

Systems biology is a discipline that is concerned with understanding the behavior of biochemical systems in terms of their parts. This includes how perturbations in external factors or how variation in protein expression levels influence system behavior. More broadly it is also concerned with understanding the operational principles and logic by which cells manage resource allocation and growth. In many respects, systems biology is similar to physiology but whereas physiology has both quantitative and qualitative aspects, systems biology is firmly quantitative in its approach. 

The easiest question to answer in systems biology is how components contribute to the overall behavior of a system. For example, consider the simple biochemical pathway:
\begin{equation}
X_o \stackrel{e_1}{\longrightarrow} x_1 \stackrel{e_2}{\longrightarrow} x_2 \stackrel{e_3}{\longrightarrow}
\label{eqn:SimplePathway}
\end{equation}
which has three reaction steps catalyzed by three enzymes $e_1, e_2$, and $e_3$, a boundary species $X_o$ which we assume to be fixed so that the system can reach a steady-state (also called the fixed point or equilibrium point in other disciplines), and two species, $x_1$ and $x_2$ which can evolve in time. If the rates of the three reactions are given by $v_1, v_2$, and $v_3$ respectively, using mass-conservation we derive the differential equations for this system as:
\begin{align}
    \begin{split}
        \frac{dx_1}{dt} = v_1 - v_2 \\[4pt]
        \frac{dx_2}{dt} = v_2 - v_3
    \end{split}
\end{align}
Note there is no differential equation for $X_o$ because $X_o$ is a boundary species that is considered an input to the system. As an input, we can in principle apply any kind of input signal to $X_o$. The most common signal, however,  is a step input, which essentially makes the input a constant. $X_o$ isn't the only possible input, the enzymes along the pathway can also be designated inputs since we can imagine up-regulating their expression. Theoretically, any parameter of the system can be designated an input, though in practice not all parameters can be manipulated experimentally.

In this article, the words parameter and input will be considered synonymous; sometimes I will use the word parameter instead of input because it will be more appropriate in those instances.

There are some obvious questions we can ask, for example, how does enzyme $e_2$ influence the steady-state concentrations of species $x_1$ and $x_2$? If we were to increase $e_2$ does that increase or decrease the steady-state level of $x_1$? In addition, what determines the degree of influence $e_2$ might have of $x_1$ and $x_2$? 

One way to answer such questions is to build quantitative models using differential equations from which we can run computer simulations to obtain time courses. However, trying to understand a system this way is not easy unless the system is very simple, and even then subtleties can be missed. For more complex systems, understanding the system's behavior by relying only on computer simulation is surprisingly difficult. Instead, various theoretical and numerical tools have been developed to assist a modeler in unraveling why a system operates the way it does. 

One of the most common methods is to focus on the steady-state and study perturbations around the steady-state. There is however one issue that limits what we can do. Most models found in systems biology, other than the simplest, are nonlinear. For example, reaction rates might be governed by nonlinear equations such as Hill equations or other enzymatic rate laws such as Michaelis-Menten rate laws. In addition, many reactions, particularly in signaling networks, are bimolecular which can also lead to nonlinearities. Nonlinearity makes such systems mathematically intractable. As a result, it is common practice to linearize\footnote{\url{https://en.wikipedia.org/wiki/Linearization}} such systems around the steady-state. This makes the mathematical analysis tractable however it does limit the analysis to regions close to the steady-state. Even with these restrictions, a wealth of knowledge can still be obtained from linearized models. This is the basis of theoretical approaches such as  metabolic control analysis and the closely related biochemical systems theory~\citep{savageau1972behavior,kacser1973control,heinrich1974linear,sauroMCABook}

To study nonlinear behavior other techniques are available such as the use of phase portraits, bifurcation analysis, and a host of methods if the system exhibits chaotic behavior. In general, however, understanding nonlinear systems is difficult. 

In this article, the focus will be on one specific method, the frequency response, that can be employed when studying linearized models.

I should say a few words on some specific terminology, $X_o$, from a control theory perspective, is called an input. In systems biology, such inputs are also called boundary species~\citep{hucka2003systems}.  Species that are not classed as inputs, such as $x$ in scheme~\eqref{eqn:SimplePathway}, are called state variables in control theory or floating species in systems biology. Floating species will evolve in time as a result of the operation of the model. In a later section, I will use the terminology `floating species' rather than `state variable' and one should be aware of their meaning. 

\section{Frequency Response}

Because the frequency response of a system is not well known by the systems biology community and surprisingly, by the dynamical systems community, it is worth explaining this technique in a little detail.

A common technique that is used in designing and understanding man-made devices is the frequency response. This technique has not found much use in systems biology~\citep{mettetal2008frequency} but is widely used in engineering fields. Operationally obtaining the frequency response involves applying a sine wave signal at a chosen input and examining how that time-varying signal is manifested at the system outputs. For example, in the pathway shown in~\eqref{eqn:SimplePathway}, we might apply a sine wave to the boundary species $X_o$ and observe how this sine wave propagates to the internal species $x_1$ and $x_2$. Intuition suggests that both $x_1$ and $x_2$ will also begin to vary in response to the input and this is indeed the case. However, the components in the system will tend to distort the sine wave and the oscillations we see in $x_1$ and $x_2$ will not exactly match the input sine wave applied to $X_o$. Of particular interest is that a linear model (or a nonlinear model that has been linearized), behaves in a special way when a sinusoidal signal is input to the system. In this situation, the system will distort the input sine wave by only changing the amplitude and phase of the sine wave. The amount that the system changes the amplitude and phase is dependent on the frequency of the input wave. Such changes can give useful information on the properties of the system. It is important to note that the frequency component of a sinusoidal signal is unchanged in a linear system.

Computing the frequency response involves computing the steady-state of the system and applying a sine wave at a given frequency and amplitude at one of the inputs. Since the input is now varying, the system responds by following the variation, however, due to delays in the system and the potential for attenuation or amplification of the sine wave, there will be changes in the amplitude and phase compared to the sinusoidal at the input. By varying the frequency of the input sine wave we can observe how the system responds across a range of frequencies. 

In practice, we don't physically inject a sine wave into a computational model but can instead be done mathematically. For a physical system such as an electronic circuit, the frequency response is often obtained by actually injecting a sine wave into the circuit.

\begin{figure}
\centering
\includegraphics[scale=0.65]{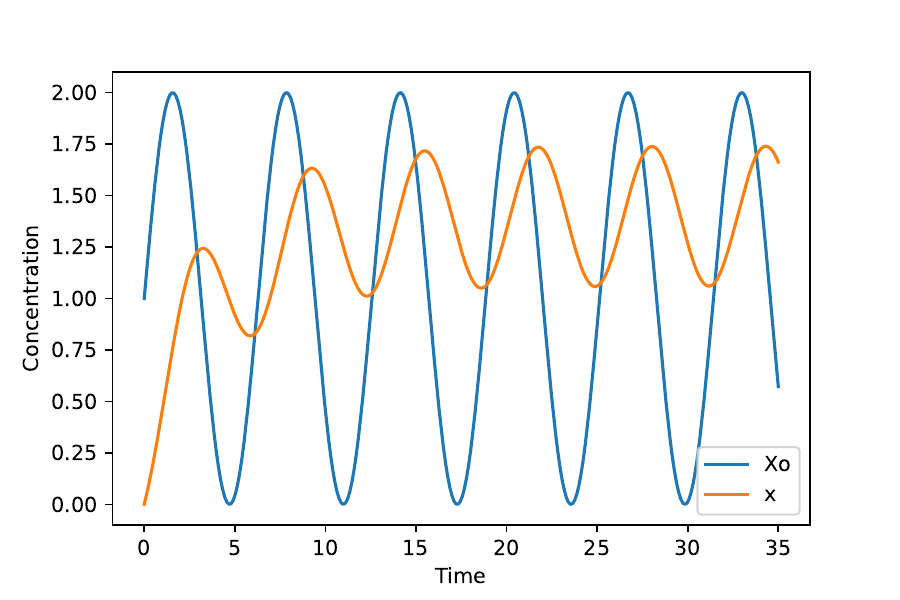}
\caption{Injecting a sine wave into a simple two-step pathway~\eqref{eqn:SimplePathway} at $X_o$. The input sine wave is shown as the blue curve, $X_o$. The intermediate concentration of $x$ is shown in the orange curve. Note that it takes a little time for $x$ to stabilize to a regular sinusoidal. Once it has we can see that its amplitude is reduced and its phase is shifted. That is the peaks don't match the peaks in the blue input sine wave. The observed phase shift is due to the time it takes to fill and empty the intermediate pool $x$.}
\end{figure}

The mathematical techniques for obtaining the frequency response are well-known in engineering fields such as electrical and mechanical engineering and it is possible to those these techniques directly. However, there is an advantage to modifying these slightly by exploiting specific biological information such as the stoichiometry matrix and the reaction step elasticities~\citep{kacser1973control, sauroMCABook}. Ingalls~\citep{ingalls2004frequency} was the first to examine the frequency response from this perspective. One advantage of this approach is that it is possible to implement algorithms for computing the frequency response for any kind of biochemical network.

In this article, I will describe a Python package that allows a user to investigate a model, either derived from an SBML~\citep{hucka2003systems} file or an antimony description~\citep{smith2009antimony} and compute the frequency response. Moreover, users can select any input and observe the frequency response on any output. This allows the user to obtain all combinations of inputs and outputs should they need to. The package uses libroadrunner~\citep{welsh2023libroadrunner} as the simulation engine. In principle, however, any equivalent simulation engine will work so long as it can compute the steady-state, the elasticities, the reduced stoichiometry matrix, and the link matrix~\citep{reder1988metabolic}. 

\section{Theory}

The frequency response of a system is based on transforming the system into the frequency domain. This assumes that the model is cast as a set of ordinary differential equations. The following description is based on the work by Ingalls~\citep{ingalls2004frequency}. An additional publication was also published shortly afterward by Rao et al~\citep{rao2004putting} who described a similar approach. The tutorial by Schulthess et al~\cite{schulthess2018frequency}, has a useful flowchart in their Figure 1 that describes some elements of the workflow. 

For convenience, the symbols and dimensions used in the discussion are shown in Table~\ref{tbl:dimensions}. The starting point for the analysis is the system equation~\citep{heinrich2012regulation,hofmeyr2001metabolic} that describes the rate of change of species in a biochemical network:
\begin{equation}
\frac{d \mathbf{x}}{d t}=\mathbf{N v}(\mathbf{x}, \mathbf{p})
\label{eqn:SystemEquation}
\end{equation}
In this equation, $\mathbf{N}$ is the $n \times r$ stoichiometry matrix containing the stoichiometric coefficients of the floating species, with $n$ being the number of floating species, and $r$ the number of reactions.  $\mathbf{v}$ is the $r$ dimensional vector of reaction rates; $\mathbf{x}$ is the $n$ dimensional vector of floating species and $\mathbf{p}$ the $p$ dimensional vector of inputs. The $\mathbf{p}$ vector will depend on the application. All these are described in Table~\ref{tbl:dimensions}.

In metabolic control analysis, the inputs are often the vector of enzyme activities. However, it could also be the vector of boundary species that marks the inputs and outputs of the pathway or it could be any parameter such as a rate constant. Obviously in practice rate constants can not be varied in a sinusoidal manner but theoretically, we could investigate what would happen if they could be varied.

\begin{table}
\centering
\begin{tabular}{ll} \toprule
Symbol & Meaning \\ \midrule
$n$ & Total number of species (not including boundary species) \\[2pt]
$n_\text{I}$ & Number of independent species \\[2pt]
$r$ & Number of reactions \\[2pt]
$p$ & Number of inputs \\[2pt]
$\mathbf{N}$ & $n \times r$ stoichiometry matrix \\[2pt]
$\mathbf{v}$ & $r$ dimension vector of rates \\[2pt]
$\mathbf{p}$ & $p$ dimension vector of inputs \\[4pt]
$\dfrac{\partial \mathbf{v}}{\partial \mathbf{x}}$ & $r \times n$ unscaled elasticity matrix \\[9pt]
$\dfrac{\partial \mathbf{v}}{\partial \mathbf{p}}$ & $r \times p$ unscaled elasticity input matrix \\[11pt]
$\mathbf{N}_r$ & $n_\text{I} \times r$ reduced stoichiometry matrix \\[5pt]
$\ \mathbf{L}$ & $n \times n_\text{I}$ link matrix \\
\bottomrule \\
\end{tabular}
\caption{Matrix symbols and dimensions}
\label{tbl:dimensions}
\end{table}

The system equation~\eqref{eqn:SystemEquation} is in general nonlinear and must therefore be linearized. When linearizing, a suitable operating point must also be chosen and the most common is the steady-state. The system is at steady-state when the left-hand side of the system equation is set to zero. Therefore setting the system equation to zero gives:
$$ \frac{d \mathbf{x}}{d t}=\mathbf{N v}(\mathbf{x}, \mathbf{p}) = 0$$
and linearizing using the Taylor series and truncation at the second term, yields:
\begin{equation}
\frac{d \delta \mathbf{x}}{d t}=\mathbf{N} \frac{\partial \mathbf{v}\left(\mathbf{x}_{s s}, \mathbf{p}\right)}{\partial \mathbf{x}} \delta \mathbf{x}+\mathbf{N} \frac{\partial \mathbf{v}\left(\mathbf{x}_{s s}, \mathbf{p}\right)}{\partial \mathbf{p}} \delta \mathbf{p}
\label{eqn:linearized1}
\end{equation}
Note, $\mathbf{N}$ is considered a constant in this operation. The subscript $ss$ is used to indicate that these values are the steady-state values.

The term $\frac{\partial \mathbf{v}\left(\mathbf{x}_{s s}, \mathbf{p}\right)}{\partial \mathbf{x}}$ is a matrix of unscaled elasticities and has dimensions: $r \times n$.  We will replace this term with the less cumbersome symbol:
$$ \frac{\partial \mathbf{v}}{\partial \mathbf{x}} $$
Likewise $\frac{\partial \mathbf{v}\left(\mathbf{x}_{s s}, \mathbf{p}\right)}{\partial \mathbf{p}}$ is a matrix of unscaled elasticities related to the inputs to the system. If there are $p$ inputs then the dimensions of this matrix will be $r \times p$. We will replace this term with the less cumbersome symbol:
$$ \frac{\partial \mathbf{v}}{\partial \mathbf{p}} $$
Using the new symbols, we can express equation~\eqref{eqn:linearized1} as:
\begin{equation}
\frac{d \delta \mathbf{x}}{d t}=\mathbf{N} \frac{\partial \mathbf{v}}{\partial \mathbf{x}} \delta \mathbf{x}+\mathbf{N} \frac{\partial \mathbf{v}}{\partial \mathbf{p}} \delta \mathbf{p}
\label{eqn:linearized2}
\end{equation}
Those familiar with control theory will realize that equation~\eqref{eqn:linearized2} is a linear time-invariant system or LTI.

To examine the frequency response we must obtain the transfer function $\mathbf{H}(s)$. To do this we move equation~\eqref{eqn:linearized2} into the Laplace domain~\citep{sauroLaplaceTransformBook}, and assuming initial conditions are zero for the left-hand side, we solve for $\mathbf{X}(s)$. Note that $d \delta \mathbf{x}/d t$ is transformed into $s \mathbf{X}(s)$. After rearrangement, and solving for $\mathbf{X}(s)$, we obtain:
\begin{equation}
\mathbf{X}(s)=\left(s \mathbf{I}-\mathbf{N} \frac{\partial \mathbf{v}}{\partial \mathbf{x}}\right)^{-1} \mathbf{N} \frac{\partial \mathbf{v}}{\partial \mathbf{p}}\ \mathbf{P}(s)
\label{eqn:XsTranform}
\end{equation}
where $\mathbf{I}$ is the $n \times n$ identity matrix and $s$ is the complex number, $a + jb$ that the Laplace transform introduces. The equation involves a matrix inversion and we will assume that such an inversion can be done. Cases where the inverse cannot be computed include the presence of conserved moieties (which we will return to) or a model that is not realistic. 

The transfer function, $H(s)$, for a specific input and output, is defined as the ratio of the transform of the output to the transform of the input~\citep{ogata2002modern}:
$$ H(s) = \frac{X(s)}{P(s)}\quad\text{or}\quad X(s) = H(s) P(s) $$
In terms of a matrix formulation, this would be written as:
$$ \mathbf{X}(s) = \mathbf{H_x}(s) \mathbf{P}(s) $$
where $\mathbf{H_x}(s)$ is the matrix form of the transfer function and includes all combinations of inputs and outputs. Comparing the above equation to~\eqref{eqn:XsTranform} we can surmise that the transfer function must be equal to:
\begin{equation}
\mathbf{H}_\text{x}(s)=\left(s \mathbf{I}-\mathbf{N} \frac{\partial \mathbf{v}}{\partial \mathbf{x}}\right)^{-1} \mathbf{N} \frac{\partial \mathbf{v}}{\partial \mathbf{p}}
\label{eqn:Hxs1}
\end{equation}
To help clarify the notation, a subscript $\text{x}$ has been added to the transfer function symbol $\mathbf{H}(s)$ to indicate that this transfer function is with respect to the species concentrations, $\mathbf{x}$.

To those more familiar with control theory, this is simply the standard form obtained using the state-space approach:
\begin{equation}
\mathbf{X}(s)=(s \mathbf{I}-\boldsymbol{A})^{-1} \boldsymbol{B}\ \mathbf{P}(s)
\label{eqn:TransferStandard}
\end{equation}
This form of the equation is found in many textbooks and the Wikipedia page\footnote{~\url{https://en.wikipedia.org/wiki/State-space_representation}} offers a good summary of this equation. Matrix $\mathbf{A}$ is the Jacobian matrix, and comparing the standard form to equation~\eqref{eqn:Hxs1} indicates that:
$$ \mathbf{A} = \mathbf{N} \frac{\partial \mathbf{v}}{\partial \mathbf{x}} $$
Before proceeding, we need to determine whether the system is stable or not. If the system is unstable then a frequency response makes no sense and no further action can be made. Stability however is easily determined by examining the eigenvalues of the Jacobian matrix. If the real part of any eigenvalue is positive then the system is unstable. 

There is, however, a small complication that needs to be considered when a biochemical pathway has conserved moieties~\citep{hofmeyr1986metabolic}. Under these conditions, some of the eigenvalues will be zero due to row dependencies in the Jacobian. These eigenvalues should be ignored. Alternatively, some simulators, such as libroadrunner, can compute a reduced Jacobian matrix. In this case, the offending eigenvalues are automatically removed. 

If a system has multiple steady-states, such as a bistable system, then there will be two sets of frequency responses, one for each steady-state. Both can be independently analyzed by picking the appropriate steady-state as the starting point.

The $\mathbf{B}$ term in equation~\eqref{eqn:TransferStandard} is often called the control or input matrix and is given by
$$ \mathbf{B} = \mathbf{N} \frac{\partial \mathbf{v}}{\partial \mathbf{p}} $$
As indicated before, these results are fairly standard in control theory though updated to use network information such as the stoichiometry matrix and the unscaled elasticity matrix. Ingalls went further, however, and took the novel approach of using the output equation as used in control theory to describe the fluxes in transform space. In control theory, the output equation is often written as:
\begin{equation}
\mathbf{y}(t)=\mathbf{C} \mathbf{x}(t)+\boldsymbol{D} \mathbf{p}(t)
\label{eqn:outputEquation}
\end{equation}
Ingalls set the output, $\mathbf{y}$, to the steady-state rates or fluxes of the pathway. We can express the reaction rate equation in vector form using:
$$ \mathbf{v}=\mathbf{v}(\mathbf{x}, \mathbf{p}) $$
where each rate is a function of species concentrations and inputs. Linearization of this equation gives:
$$\delta \mathbf{v}=\frac{\partial \mathbf{v}}{\partial \mathbf{x}} \delta \mathbf{x}+\frac{\partial \mathbf{v}}{\partial \mathbf{p}} \delta \mathbf{p}
$$
This now looks like the output equation~\eqref{eqn:outputEquation}. Taking Laplace transforms on both sides gives:
$$ \boldsymbol{V}(s)=\frac{\partial \mathbf{v}}{\partial \mathbf{x}}  \boldsymbol{X}(s)+ \frac{\partial \mathbf{v}}{\partial \mathbf{p}}  \boldsymbol{P}(s) $$
Note that the transfer function for $\mathbf{X}(s)$ appears in this equation. This means we can substitute equation~\eqref{eqn:XsTranform} into this equation to give:
$$ \boldsymbol{V}(s) = \frac{\partial \mathbf{v}}{\partial \mathbf{x}} 
\left(s \mathbf{I}-\mathbf{N} \frac{\partial \mathbf{v}}{\partial \mathbf{x}}\right)^{-1} \mathbf{N} \frac{\partial \mathbf{v}}{\partial \mathbf{p}}\ \mathbf{P}(s) + \frac{\partial \mathbf{v}}{\partial \mathbf{p}}  \boldsymbol{P}(s)
$$
$\mathbf{P}(s)$ can be moved to the right to give:
$$ \boldsymbol{V}(s) = \left( \frac{\partial \mathbf{v}}{\partial \mathbf{x}} 
\left(s \mathbf{I}-\mathbf{N} \frac{\partial \mathbf{v}}{\partial \mathbf{x}}\right)^{-1} \mathbf{N} \frac{\partial \mathbf{v}}{\partial \mathbf{p}} + \frac{\partial \mathbf{v}}{\partial \mathbf{p}}  \right) \boldsymbol{P}(s) = \mathbf{H}_v(s) \mathbf{P}(s)
$$
The transfer function is therefore:
$$ \mathbf{H}_v(s)=\frac{\partial \mathbf{v}}{\partial \mathbf{x}}\left(s \mathbf{I}-\mathbf{N} \frac{\partial \mathbf{v}}{\partial \mathbf{x}}\right)^{-1} \mathbf{N} \frac{\partial \mathbf{v}}{\partial \mathbf{p}}+\frac{\partial \mathbf{v}}{\partial \mathbf{p}} $$
A subscript, $v$, has been added to indicate that this is the transfer function for the fluxes in the pathway. 

To examine the frequency response we replace the complex number $s$ with $j\omega$\footnote{Out of habit and due to my more engineering background, I tend to use $j$ instead of $i$ for the imaginary number.} to move us into the Fourier domain, which gives us
\begin{align}
 \mathbf{H}_x(j\omega) &=\left(j\omega \mathbf{I}-\mathbf{N} \frac{\partial \mathbf{v}}{\partial \mathbf{x}}\right)^{-1} \mathbf{N} \frac{\partial \mathbf{v}}{\partial \mathbf{p}} 
 \label{eqn:Hxs2}
 \end{align}

 \begin{align}
\mathbf{H}_v(j\omega) &=\frac{\partial \mathbf{v}}{\partial \mathbf{x}}\left(j\omega \mathbf{I}-\mathbf{N} \frac{\partial \mathbf{v}}{\partial \mathbf{x}}\right)^{-1} \mathbf{N} \frac{\partial \mathbf{v}}{\partial \mathbf{p}}+\frac{\partial \mathbf{v}}{\partial \mathbf{p}} 
\label{eqn:Hvs2}
\end{align}

Both equations yield complex numbers and the frequency response is encoded in these values. The phase and amplitude of the frequency response can be obtained from the corresponding phase and amplitude of the complex number. Since the frequency is given by $\omega$, this equation can be used to determine the frequency response as a function of frequency. 

Classically, the frequency response is plotted in the form of Bode plots~\citep{ogata2002modern} which consists of two separate plots. One plot is used to indicate changes in the amplitude and a second is used to indicate changes in the phase. The $x$-axis in both plots marks the input frequency on a log scale which can be either expressed in radians/sec or in Hz. The amplitude plot is plotted using decibels (dB) - essentially a log scale - and the phase is plotted in degrees on a linear scale.

\subsection*{Generalization}

Some biochemical networks, particularly signaling networks but also metabolic systems, will contain conserved moieties. This can result in numerical issues when computing the inverse because the Jacobian becomes singular. Equations~\eqref{eqn:Hxs2} and~\eqref{eqn:Hvs2} must be adjusted to include the link matrix~\citep{reder1988metabolic}, $\mathbf{L},$ and the reduced stoichiometry matrix~\citep{reder1988metabolic}, $\mathbf{N}_r$. Details of this can be found in the Ingalls paper~\citep{ingalls2004frequency} although equation (9) in that paper has an error where the $\mathbf{L}$ matrix has been inadvertently left out. Note that Ingalls also only gives the frequency response equation for the independent species he calls $s_i$. The full set can be obtained by premultiplying by $\mathbf{L}$~\citep{hofmeyr2001metabolic}. The complete set of generalized equations is shown in~\eqref{eqn:HxLink}.

\begin{align}
\begin{split}
\mathbf{H}_{x}(j \omega) = \mathbf{L}
\left(j \omega \mathbf{I}-\mathbf{N}_{\mathrm{R}} \frac{\partial \mathbf{v}}{\partial \mathbf{s}} \mathbf{L}\right)^{-1} 
\mathbf{N}_{\mathrm{R}} \frac{\partial \mathbf{v}}{\partial \mathbf{p}} \\
\mathbf{H}_v(j \omega)= \mathbf{L} \frac{\partial \mathbf{v}}{\partial \mathbf{s}}\left(j \omega \mathbf{I}-\mathbf{N}_{\mathrm{R}} \frac{\partial \mathbf{v}}{\partial \mathbf{s}} \mathbf{L}\right)^{-1} \mathbf{N}_{\mathrm{R}} \frac{\partial \mathbf{v}}{\partial \mathbf{p}}+\frac{\partial \mathbf{v}}{\partial \mathbf{p}} 
\end{split}
\label{eqn:HxLink}
\end{align}

These are the equations we will use to compute the frequency response. 

Ingalls also highlighted the important fact that at zero frequency, that is $\mathbf{H}(0)$, equations~\eqref{eqn:HxLink} reduce to the unscaled control coefficients found in metabolic control analysis~\citep{kacser1995control,heinrich2012regulation,hofmeyr2001metabolic,sauroMCABook}. This result showed that MCA and classical control theory were one and the same thing though MCA derived additional theorems due to the stoichiometric nature of biochemical networks.

\section{Example}

To illustrate the use of these equations let us consider a simple two-step pathway:
$$ X_o \stackrel{v_1}{\longrightarrow} x \stackrel{v_2}{\longrightarrow} $$
and compute the species frequency response using equation~\eqref{eqn:Hxs2}. We can use equation~\eqref{eqn:Hxs2} in this example because the model does not contain any conserved moieties. 

To keep things simple, we will assume that $v_1 = k_1 X_o$ and $v_2 = k_2 x$. This is a one-dimensional system with a single variable $x$ but we will assume two inputs, $k_1$ and $k_2$ to make it more interesting.  The differential equation for this system is:
$$ \frac{dx}{dt} = k_1 X_o - k_2 x $$
We can define the sizes for the various components using $n = 1$, $r = 2$, and $p = 2$. Matrix $\mathbf{A}$ is therefore a simple $1 \times 1$ matrix. The $\mathbf{B}$ matrix however is a $2 \times 2$ matrix. But as we'll see only the main diagonal has entries because $k_1$ has no direct effect on the second reaction and $k_2$ has no direct effect on the first reaction.

The $\mathbf{A}$ matrix is given by:
$$ \mathbf{A} =  \mathbf{N} \frac{\partial \mathbf{v}}{\partial \mathbf{x}} 
= \begin{bmatrix}
    1 & -1
\end{bmatrix}
\begin{bmatrix}
    \dfrac{\partial v_1}{\partial x} \\[10pt]
    \dfrac{\partial v_2}{\partial x}
\end{bmatrix} 
$$
Noting that $\partial v_1/ \partial x = 0$ and $\partial v_2 /\partial x = k_2$ we obtain:
$$
\mathbf{A} = \begin{bmatrix} 
 -k_2 
 \end{bmatrix}
 $$
The eigenvalue for a 1 by 1 matrix is the entry itself. The eigenvalue is therefore $-k_2$ which is negative therefore the system is stable.

The $\mathbf{B}$ matrix can be computed in a similar manner, except that $\partial \mathbf{v}/\partial \mathbf{p}$ is a $2 \times 2$ matrix because there are two reactions and two possible inputs:
$$ \mathbf{B} = 
\mathbf{N} \frac{\partial \mathbf{v}}{\partial \mathbf{p}} = 
\begin{bmatrix}
    1 & -1
\end{bmatrix}
\begin{bmatrix}
    X_o & 0 \\
    0 & x
\end{bmatrix}
= 
\begin{bmatrix}
    X_o & -x
\end{bmatrix}
$$
To compute the frequency response for the system we need to insert these elements into equations~\eqref{eqn:Hxs2} and~\eqref{eqn:Hvs2}. As an example, let's consider deriving $\mathbf{H}_\text{x}(j\omega)$:
\begin{align}
 \mathbf{H}_\text{x}(j\omega) &=\left(j\omega \mathbf{I}-\mathbf{N} \frac{\partial \mathbf{v}}{\partial \mathbf{x}}\right)^{-1} \mathbf{N} \frac{\partial \mathbf{v}}{\partial \mathbf{p}} 
 \label{eqn:Hxs}
 \end{align}
Inserting the various terms gives:
$$ \mathbf{H}_x(j\omega) = \left(j\omega - (-k_2) \right)^{-1}
\begin{bmatrix}
X_o & -x
\end{bmatrix} = 
\begin{bmatrix}
    \dfrac{X_o}{j\omega + k_2} & \dfrac{-x}{j\omega + k_2}
\end{bmatrix}
$$
The solution has two transfer functions, one related to how changes in $k_1$ affect $x$ and a second on how changes in $k_2$ affect $x$.

The frequency response can be computed by examining the magnitude and phase components of the complex numbers $X_o/(j\omega + k_2)$ and $-x/(j\omega + k_2)$. These complex numbers need to be rationalized first into the form $a + jb$, from which the amplitude can be computed using the Pythagoras theorem:
$$ \text{Amplitude}_{k_1} = \frac{X_o}{\sqrt{k_2^2+\omega^2}}, \quad \text{Amplitude}_{k_2} = \frac{x}{\sqrt{k_2^2+\omega^2}} $$
and the phase changes using basic trig are:
$$ \text{Phase}_{k_1} = \tan ^{-1}\left(\frac{-\omega}{k_2}\right),\quad \text{Phase}_{k_2} = \tan ^{-1}\left(\frac{\omega}{-k_2}\right) $$
The phase depends on what quadrant we are in on the complex plane. The phase shift, with respect to $k_1$, with vary from 0 to $-90$ degrees, indicating that the phase shift lags the signal. For the phase shift with respect to $k_2$, there is a difference. This time the phase shift starts at $+180$ degrees and decreases to $+90$ degrees as the frequency increases. Note that the phase shift is positive for input $k_2$ which might suggest $x$ is anticipating changes in $k_2$. This is an illusion, because when $k_2$ increases, consumption of $x$ increases so that $x$ decreases meaning the change in $x$ is opposite to the change in $k_2$ but with a delay in the response.

This example is illustrated in a later section where we demonstrate the software. 

As we've seen, for small systems, it is possible to derive the analytical expressions for the frequency response but one can imagine that for large systems this can become quite tedious. The purpose of the Python package is to compute the amplitude and phase changes numerically for any sized network, with any number of inputs and outputs. 

\section{Python Implementation}

Python is a good language to evaluate equations such as~\eqref{eqn:HxLink} because matrix multiplication and other manipulations are supported by the {\tt numpy} library and Python natively supports complex numbers.  As an illustration, part of the computation is shown in Listing 1.

The code begins by ensuring the system is at steady-state. If it can't find a steady-state the method throws an exception. libroadrunner has a specific method for computing steady-state as do other simulators used by the systems biology community. The variable {\tt r} is a libroadrunner instance but could be another simulator that supports the features we need. 

There is one minor complication when constructing the matrices. Because everything has to be computed using complex numbers we have to ensure that matrices such as the link matrix or stoichiometry matrix are converted to complex types. Moreover, libroadrunner returns named matrices so they must also be converted to pure {\tt numpy} types using the {\tt numpy} method {\tt array()}.

The evaluation of the matrix equation is fairly straightforward when using the {\tt numpy} library. The code shows the individual steps as the matrix expression is constructed piece by piece. This was done to make reading the code easier and was also useful during the debugging process. 

Another minor complication was computing the $\partial \mathbf{v}/\partial \mathbf{p}$ matrix. libroadrunner does not have a single method that can return the full matrix. The matrix was therefore constructed using a loop by calling the libroadrunner method {\tt getUnscaledParameterElasticity} for each reaction. The reader may also notice that even though the unscaled parameter elasticity matrix should be a $r \times p$ matrix, the code appears to only construct a vector of size $r$.  The reason for this is that the code as written will only compute the transfer function for a specific input. Hence we need only consider a column of $\partial \mathbf{v}/\partial \mathbf{p}$ that corresponds to that input. This does not preclude the possibility of looking at all possible inputs, it just means that code needs to be called multiple times with each input. Once the transfer function has been computed we use Python functions {\tt abs} and {\tt math.atan2} to extract the amplitude and phase respectively. The {\tt atan2} function takes into account the quadrant to compute the correct value for the degrees. 

The code is packaged into a Python function to make it easy to use. The function will compute the frequency response for a given output with respect to a given input. Two methods are provided, one for computing the species concentration frequency response and another for computing the rate or flux frequency response. 

As an example, the species frequency response method has the signature:

{\tt def getSpeciesFrequencyResponse(startFrequency, numberOfDecades, numberOfPoints, parameterId, variableId, useDB=True, useHz = True)}

The input is specified by the string argument {\tt parameterId} and the output by the string argument {\tt variableId}. To get the full set of transfer functions a user should call this method repeatedly with changes to these two arguments to get the full set of frequency response combinations. This does incur some inefficiency since virtually the same computation is done each time but it is often the case that only specific transfer functions are of interest. A future version might add to the code the ability to request all transfer functions to make the code more efficient. 

One final aspect of the code that might be of interest is a small complication when generating the Bode phase plots. When scanning over a range of frequencies the phase shift can often go beyond -180 degrees. Normally this will result in a wraparound back to +180 degrees which when plotted appears as a discontinuity in the plot. To avoid this and to ensure a continuous appearance in the phase, the {\tt numpy.unwrap} function is used to unwrap a signal at the discontinuous point which by default happens to be at multiples 180 degrees. Thus when the phase passes -180 degrees, the function will ensure that the phase continues beyond -180 rather than jumping back to +180.  

\medskip
\begin{lstlisting}[caption={Implementation of part of the computation}]
try:
   r.steadyState()
except Exception as e:
   raise Exception ("There doesn't appear to be a steady state");
   
linkMatrix = r.getLinkMatrix()
linkMatrix = np.array (linkMatrix)
linkMatrix = linkMatrix.astype (complex)

ee = r.getUnscaledElasticityMatrix()
ee = np.array (ee)
ee = ee.astype (complex)
        
# Convert stoichiometry matrix to complex matrix
Nr = r.getReducedStoichiometryMatrix()
Nr = np.array (Nr)
Nr = Nr.astype (complex)

# Compute the Jacobian = Nr dv/ds L
Jac = np.matmul (Nr, ee)
Jac = np.matmul (Jac, linkMatrix)
identMatrix = np.identity(numFloatingSpecies, dtype=complex)

# Compute Hx(jw) = L (jw I - Nr dv/dx L )^{-1} Nr dv/dp
jw_val = complex (0, wf)
T1 = identMatrix*jw_val         # = jwI
T2 = T1 - Jac                   # = jwI - Jac
inverse = np.linalg.inv(T2)     # = (jwI - Jac)^(-1)
T3 = np.matmul(inverse, Nr)     # = (jwI - Jac)^(-1) Nr
T3 = np.matmul(linkMatrix, T3)  # = L (Nr dv/ds L)^(-1) Nr

dvdp = np.zeros(numReactions, dtype=complex)
for k in range (numReactions): 
    val = r.getUnscaledParameterElasticity(reactionIds[k], parameterId);
          dvdp[k] = val+0j
T4 = np.matmul (T3, dvdp) 
\end{lstlisting}

\section{Illustrations}

With the theory and code described it is worth looking at some applications. First, we'll describe how to use the code.

\subsection*{Obtaining and Using the Code}

The code can be obtained from~\url{https://github.com/sys-bio/frequencyResponse}. There is no python pip installer because the code is just a single file called {\tt freqResponse.py}. However, it is important to {\tt pip install tellurium} because this is needed to supply the {\tt libroadrunner} simulator and antimony support. 

Once you download the {\tt freqResponse.py} file, copy it to a location where
you plan to use it. After that you need to load your model into libroadrunner. The code below illustrates this:

\medskip
\begin{lstlisting}[caption={Loading a model into libroadrunner using Tellurium}]
import tellurium as te
from freqResponse import *

r = te.loada('''
    $Xo -> x; k1*Xo
      x -> ;  k2*x

     k1 = 0.1; k2 = 0.34
     Xo = 4
''')
\end{lstlisting}

where I've used the antimony syntax for describing a reaction network, but it could equally be an SBML based model. The code also imports {\tt freqResponse} to make available all the methods in that file. In all the presented examples I am not going to include any explicit sink species. This is why the last reaction has an empty right-hand side. A missing sink is not important in these examples but can be included depending on the application.

The call to {\tt loada} returns a roadrunner instance, {\tt r}. This is required by the frequency response code.

To compute a frequency response, first, pass the libroadrunner instance to the function {\tt FrequencyResponse} as follows;

{\tt fr = FrequencyResponse (r)}

This will return a frequency response object, which I have called {\tt fr}. Using {\tt fr} one can request the frequency response with respect to species concentrations or fluxes. For example, to compute a species frequency response we call the following method:

{\tt results = fr.getSpeciesFrequencyResponse(0.01, 3, 1000, 'X0', 'x3')}

The arguments include the starting frequency (0.01), the number of decades on the frequency x-axis (3) from the starting frequency, the number of points to generate (1000), the symbol name of the input from the model (`Xo'), and the symbol name of the output species (`x3') we wish to look at. The method for the flux frequency response has the same argument list except the output symbol must be a reaction name. In theory, I could have merged the two functions into one and the method could have recognized whether it was dealing with a species or a reaction flux. There are other optional arguments such as using Hz or rad/sec for the frequency units.

There is also a convenience method called {\tt plot} which will plot the Bode plots based on the last call. The user also has access to the raw data through the results returned from the call. This will contain three columns, frequency, amplitude, and phase.

\subsection*{Illustrative examples}

The simplest example we can try is a linear chain of reactions. It doesn't matter what the kinetic laws are because these will be automatically linearized in the algorithm. This means we can use Michaelis-Menten equations if we so wish. For illustration, however, let's just use simple mass-action kinetics. To begin we define the model using antimony (or a preexisting SBML file can also be used). A simple two-step pathway is shown below expressed as a string in antimony format:

\medskip
\begin{lstlisting}[caption={Antimony model of a two-step pathway}]
model = '''
    J1: $Xo -> S1; k1*Xo
    J2:   x -> ;   k2*x
    
    k1 = 0.1; k2 = 0.23; 
    Xo = 1
'''
\end{lstlisting}

To compute the frequency response of $x$ with respect to input $X_o$ we use the method calls shown below. The package also provides a convenience method, {\tt plot}, for plotting the results with an optional argument to generate pdf output for publication purposes.

\bigskip\bigskip\bigskip
\begin{lstlisting}[caption={Compute the frequency response of $x$ with respect to the input $X_o$}]
import tellurium as te
from freqResponse import *

# load the model into libroadrunner
r = r.loada (model)
# pass the libroadrunner instance to the frequency response package
fr = FrequencyResponse(r)
# compute the species frequency response
results = fr.getSpeciesFrequencyResponse(0.01, 3, 100, 'Xo', 'x')
# plot the resulting Bode plots
fr.plot()
\end{lstlisting}

\begin{figure}
\centering
\includegraphics[scale=0.75]{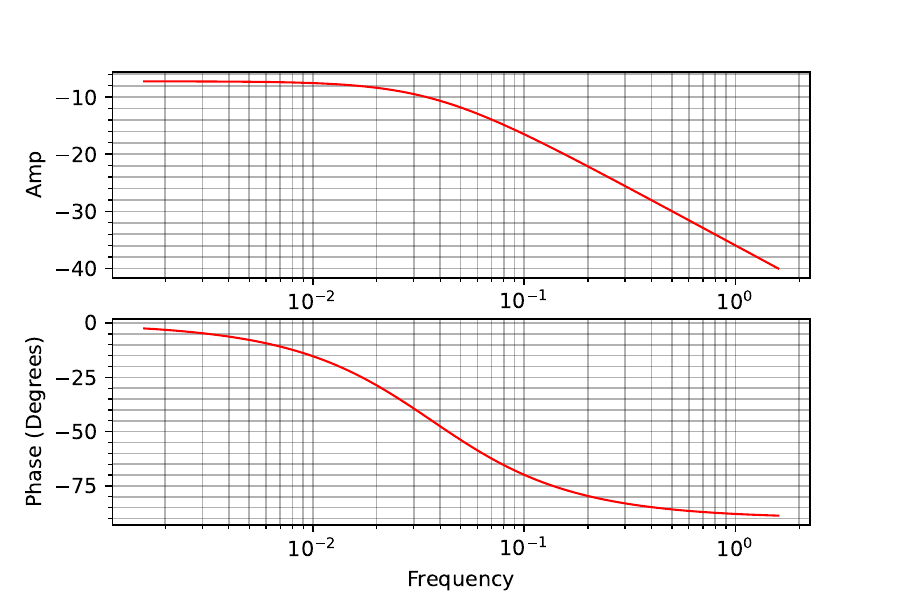}
\caption{Frequency response for a two-step linear chain with respect to the intermediate $x$ given the input $X_o$.}
\label{fig:freqRespTwoStep}
\end{figure}

Running this code yields the plots shown in Figure~\ref{fig:freqRespTwoStep}.

The plots illustrate two characteristics of this system, technically referred to as a first-order system because it only has one variable, $x$. The first is that the upper amplitude plot shows a system that behaves as a low-pass filter. That is, at low frequencies the amplitude is not changed significantly but as the frequency increases, the amplitude decreases exponentially because the system is unable to keep up with the rapid changes at the input. Since noise is usually considered high frequency, such as system could be used to smooth out a noisy signal. 

The second observation is that the phase shift plot shows that $x$ lags the input signal. At low frequencies, the phase shift is close to zero because the system can easily adjust to the changing input. This means that the input sine wave and the resulting sine wave we see at $x$, match each other. However, as the frequency increases the system is less able to keep up, and changes in $x$ start to lag the input sine wave. The maximum phase change however has a limit for a first-order system. This limit is a phase shift of exactly -90 degrees. 

This limit to the phase shift might be counter-intuitive because one might have expected the phase to continuously shift more and more negative as the frequency increases. However, equation~\eqref{eqn:phaseShift90} confirms that the phase-shift tends to -90 degrees at high frequencies:
\begin{equation}
\text{Phase} = -\tan ^{-1}\left(\frac{\omega}{k_2}\right)
\label{eqn:phaseShift90}
\end{equation}
Why is this? Expression~\eqref{eqn:phaseShift90} shows that the smaller $k_2$ the more likely the phase shift will be -90 degrees. If $k_2$ is very small then we can assume that the rate of degradation is small. This means that the change in concentration of $x$ is dominated by the input sine wave. The maximum rate of increase in $x$ is when the input sine wave is at its maximum peak. As the input sine wave decreases the rate of increase in $x$ slows until the input sine wave crosses its inflection point. Once the input sine wave reaches the inflection point, the level of $x$ also peaks. Thus the input sine wave and the concentration of $x$ will be 90 degrees out of phase with respect to the concentration of $X_o$. This means that the delay can never be more than -90 degrees. The rate at which the phase shift approaches -90 degrees is determined by the value of the degradation constant $k_2$.

\begin{figure}
\centering
    \includegraphics[scale=0.75]{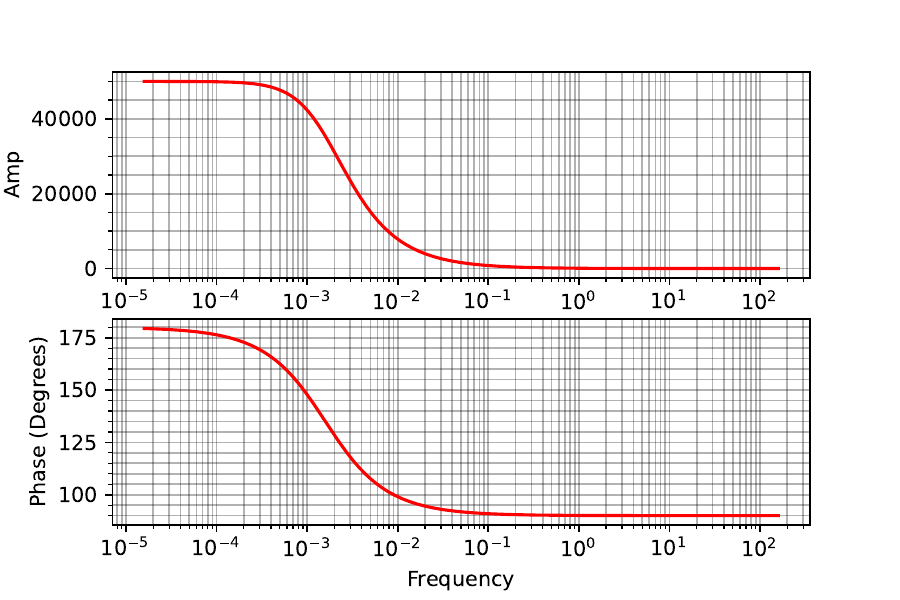}
\caption{The frequency response of a two-step pathway when the input is on the second step. Of note is that the phase-shift starts at +180 and decreases to +90 degrees as the frequency increases. This is because the input $k_2$ affects the consumption of $x$.}
\label{fig:SecondStep}
\end{figure}

In a previous section, it was also shown what the frequency response would be if the input were on the second step. The results are shown in Figure~\ref{fig:SecondStep} which confirms what we saw with the analytical solution. 

We can extend the pathway to more steps, such a model is shown below. In this case, the names of the three species are $x_1, x_2$, and $x_3$.

\medskip
\begin{lstlisting}[caption={Antimony model of a flour-step pathway}]
    J1: $Xo -> x1; k1*Xo
    J2:  x1 -> x2; k2*x1
    J3:  x2 -> x3; k3*x2
    J4;  x3 -> ;   k4*x3
    
    k1 = 0.1; k2 = 0.23 
    k3 = 0.4; k4 = 0.33
    Xo = 1
\end{lstlisting}

\begin{figure}
\centering
\includegraphics[scale=0.5]{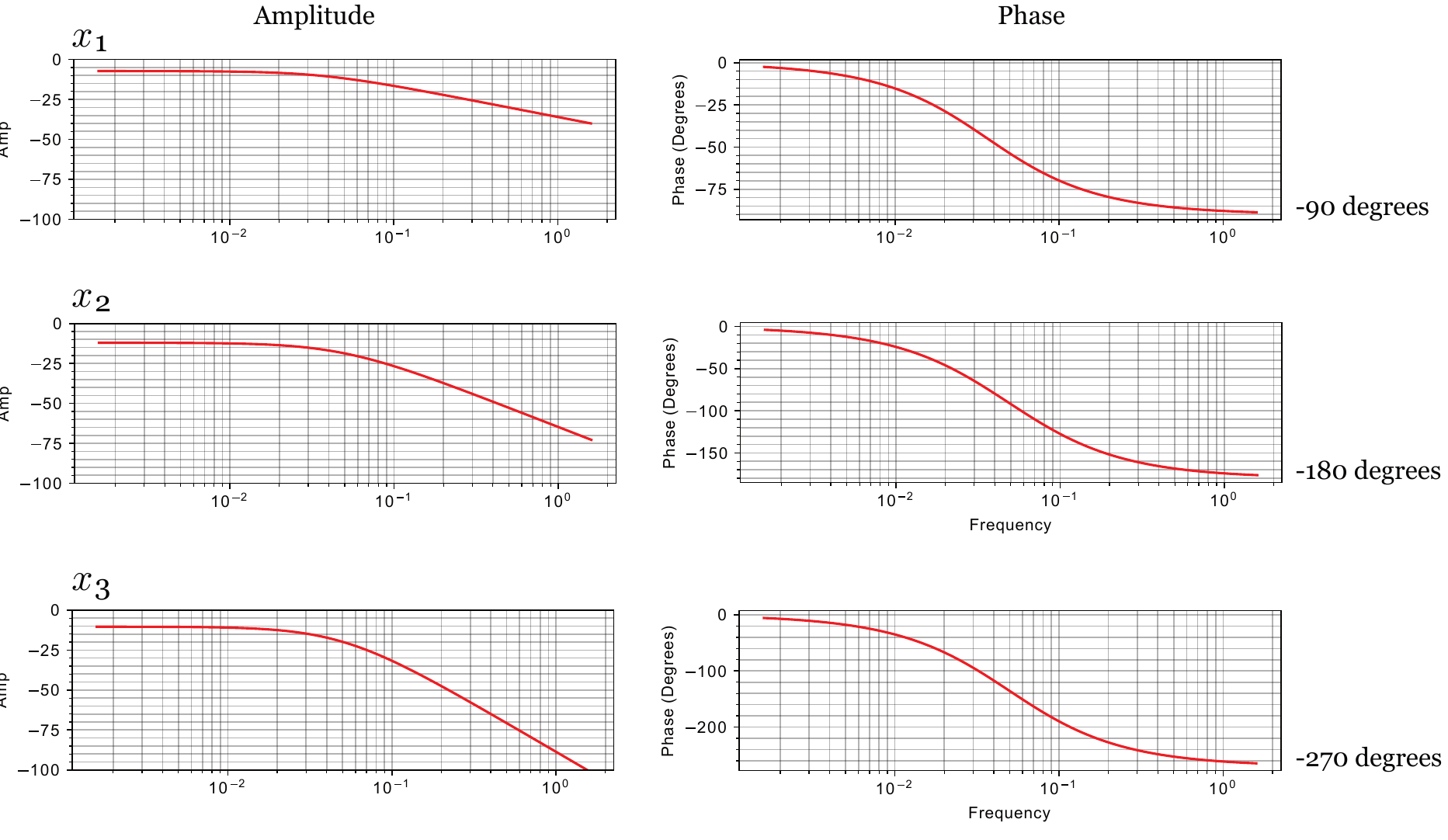}
\caption{Frequency responses for a four-step linear chain. The y-axis on amplitude plots has been fixed in all three cases (left column) so that the trends can be more easily seen. The amplitude drops as we move down the pathway from one species to the next. This is due to attenuation at each step. The phase plots on the right show the absolute value of the phase shift increasing by a maximum of -90 degrees across each species. This has important implications when coupled with negative feedback.}
\label{fig:freqResposneFourSteps}
\end{figure}

Each additional species incurs an additional maximum shift of -90 degrees. Thus $x_2$ could, under the right conditions, experience a maximum shift of -180 degrees. The third species, $x_3$ could experience an additional -90 degrees so that it could see a maximum shift of -270 degrees relative to the input $X_o$. If you look at the right panel of Figure~\ref{fig:freqResposneFourSteps} we can see that as we move down the column, the signal becomes more and more delayed.  

\subsection*{Pathway with Negative Feedback}

More interesting effects can be seen when we include negative feedback. We can adjust the four-step model as shown in Listing 6. This is a linear pathway of four steps where the last species, $x_3$ inhibits the first reaction. This forms a simple negative feedback loop. The strength of the feedback can be changed by the Hill-like coefficient $h$. This system has the potential to become unstable when $h$ is at a threshold of about 8~\citep{walter1969oscillations,savageau1972behavior}. At this point, the system will destabilize and show sustained oscillations. To avoid this we will set $h=6$ where the system is stable but still shows interesting frequency response behavior. 

\medskip
\begin{lstlisting}[caption={Antimony model of a system of four steps with negative feedback from $x_3$ to the first step.}]
  J0: $Xo -> x1; (Vm1 * Xo)/(1 + Xo + x3^h);
  J1: x1 -> x2;  k2*x1
  J2: x2 -> x3;  k3*x2
  J4: x3 -> ;    k4*x3

  x1 = 0; x2 = 0; x3 = 0;
  x4 = 0; Xo = 10; 
  k2 = 0.1; k3 = 0.11; k4 = 0.14

  Vm1 = 10; h = 6 
  V4 = 2.5; KS4 = 0.5;
\end{lstlisting}

\begin{figure}[htpb]
\centering
\includegraphics[scale=0.75]{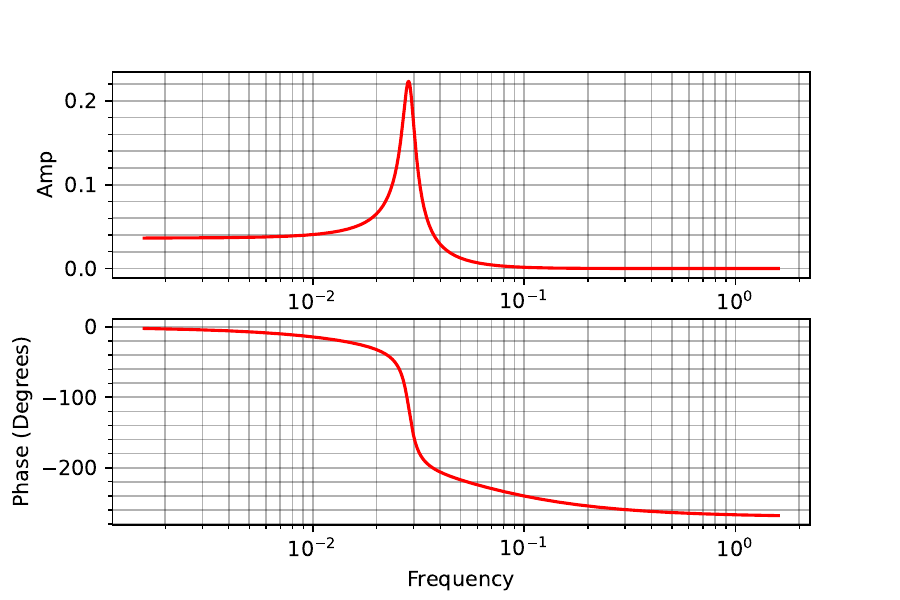}
\caption{Frequency response for a four-step linear chain with a negative feedback loop. The upper amplitude plot shows a resonance spike between a frequency of 0.01 and 0.1.}
\label{fig:FreqNegFeedback}
\end{figure}

A frequency response of this system yields the Bode plots shown in Figure~\ref{fig:FreqNegFeedback}. The biggest change can be seen in the amplitude plot where we see a sharp spike at a frequency between 0.01 and 0.1 where. This spike is a resonance spike. As the feedback strength gets stronger this spike becomes more pronounced until eventually the system destabilizes and switches to an oscillatory mode. This is a well-known phenomenon in control engineering and engineers try to design systems where the resonance spike is at frequencies that the system may not experience. An alternative strategy is to reduce the strength of the feedback but this may degrade the performance of the system at lower frequencies. 

Notice that the phase plot for $x_3$ passes through a shift of -180 degrees. This means at a shift of -180, $x_3$ changes in exactly the opposite direction to the input $X_o$ signal. This is the delay component that many authors some-what vaguely speak of when trying to explain the onset of oscillations in a feedback loop. 

The delay itself, however, is insufficient to generate oscillations. The negative feedback loop itself will also contribute a 180 degree change in the signal. The total shift around the loop is therefore 360 degrees. This means that instead of stabilizing the system, the feedback loop coupled to the delay acts as a positive feedback loop. This is what provides the destabilizing force. If at the same time, the amplitude at the -180 crossing is one or above, the system will amplify the destabilization. This ultimately results in oscillations.  A frequency response therefore gives a nice rationalization as to the origins of oscillations in a feedback loop at least for a simple feedback oscillator. Relaxation oscillators operate on a different mechanism~\citep{sauroPathwayModelingBook}.

\section*{Conclusion}

This article describes the theory, code, and some examples of using the software to compute the frequency response of any biochemical network. The last example in particular showed how an understanding of the frequency response in a system with negative feedback can explain the onset of oscillations. 

\section*{Acknowledgments}

This work was supported by the National Institutes of Health under award number U01CA227544 and by the Department of Energy award number DESC0023091. The content expressed here is solely the responsibility of the authors and does not necessarily represent the official views of the National Institutes of Health, the Department of Energy, or the University of Washington. I am also grateful for the useful discussions I had with Brian Ingalls (Waterloo), John Doyle (Caltech) and Joe Hellerstein at UW 

\bibliographystyle{apsr}
\bibliography{references}

\end{document}